\documentclass[final,5p,times,twocolumn]{elsarticle}
\usepackage{float}
\usepackage{multirow}
\usepackage{longtable}
\usepackage{booktabs}
\usepackage{amsmath}
\usepackage{threeparttable}
\usepackage{upgreek}
 \usepackage{graphicx}

\usepackage{amssymb}
\usepackage{amsthm}
\usepackage{caption}
\usepackage[nodots]{numcompress}

\usepackage{wrapfig}
\biboptions{sort&compress}
\journal{Science Bulletin}
\begin{document}
\captionsetup[figure]{labelfont={bf},labelformat={default},labelsep=period,name={Fig.}}
\captionsetup[table]{labelfont={bf},labelformat={default},labelsep=period,name={Table}}
\begin{frontmatter}



\title{Understanding angle-resolved polarized Raman scattering from black phosphorus at normal and oblique laser incidences}

\author[a]{Miao-Ling Lin\fnref{1}}
\author[a,b]{Yu-Chen Leng\fnref{1}}
\author[a,b]{Xin Cong}
\author[a,b]{Da Meng}
\author[c]{Jiahong Wang}
\author[d]{Xiao-Li Li}
\author[c]{Binlu Yu}
\author[a]{Xue-Lu Liu}
\author[c]{Xue-Feng Yu}
\author[a,b,e]{Ping-Heng Tan\corref{cor1}}
\ead{phtan@semi.ac.cn}
\cortext[cor1]{Corresponding author.}
\address[a]{State Key Laboratory of Superlattices and Microstructures, Institute of Semiconductors, Chinese Academy of Sciences, Beijing 100083, China}
\address[b]{Center of Materials Science and Optoelectronics Engineering \& CAS Center of Excellence in Topological Quantum Computation, University of Chinese Academy of Sciences, Beijing 100049, China}
\address[c]{Shenzhen Engineering Center for the Fabrication of Two-Dimensional Atomic Crystals, Shenzhen Institutes of Advanced Technology, Chinese Academy of Sciences, Shenzhen 518055, P. R. China}
\address[d]{College of Physics Science and Technology, Hebei University, Baoding 071002, China}
\address[e]{Beijing Academy of Quantum Information Science, Beijing 100193, China}
\fntext[1]{These authors contributed equally to this work}

\begin{abstract}
The selection rule for angle-resolved polarized Raman (ARPR) intensity of phonons from standard group-theoretical method in isotropic materials would break down in anisotropic layered materials (ALMs) due to birefringence and linear dichroism effects. The two effects result in depth-dependent polarization and intensity of incident laser and scattered signal inside ALMs and thus make a challenge to predict ARPR intensity at any laser incidence direction. Herein, taking in-plane anisotropic black phosphorus as a prototype, we developed a so-called birefringence-linear-dichroism (BLD) model to quantitatively understand its ARPR intensity at both normal and oblique laser incidences by the same set of real Raman tensors for certain laser excitation. No fitting parameter is needed, once the birefringence and linear dichroism effects are considered with the complex refractive indexes. An approach was proposed to experimentally determine real Raman tensor and complex refractive indexes, respectively, from the relative Raman intensity along its principle axes and incident-angle resolved reflectivity by Fresnel$'$s law. The results suggest that the previously reported ARPR intensity of ultrathin ALM flakes deposited on a multilayered substrate at normal laser incidence can be also understood based on the BLD model by considering the depth-dependent polarization and intensity of incident laser and scattered Raman signal induced by both birefringence and linear dichroism effects within ALM flakes and the interference effects in the multilayered structures, which are dependent on the excitation wavelength, thickness of ALM flakes and dielectric layers of the substrate. This work can be generally applicable to any opaque anisotropic crystals, offering a promising route to predict and manipulate the polarized behaviors of related phonons.
\end{abstract}


\end{frontmatter}


\section{Introduction}
\label{}




In-plane anisotropic layered materials (ALMs) have attracted much interest owing to their polarization-dependent optical/opto-electronic properties, leading to overwhelming potential in practical applications, such as polarization photodetectors and sensors \cite{yuan-nn-2015,zhao-afm-2018,wang-nc-2019,zhao-nadv-2020}. Through light-matter interactions, the polarization-dependent Raman intensity can provide abundant information of the lattice symmetry\cite{Loudon-book-1964,Tan-2019-booktitle,li-nmat-2019}, crystalline orientation \cite{oleksii-nc-2019} and edge features \cite{Ribeiro-nc-2016,Chai-SciAdv-2018}. For anisotropic crystals, it has been known for many years that the birefringence effect rising from anisotropic refractive index should be considered for polarized Raman intensity \cite{porto-pr-1966,Asawa-pr-1968,DAWSON-sa-1972,RULMONT-sa-1979,RULMONT-sa2-1979,Alonso-prb-2005}. In transparent anisotropic crystals, the anomalous polarized Raman intensity was attempted by introducing a phase delay of incident polarization between two principle axes and integration over the length of crystal \cite{RULMONT-sa-1979,RULMONT-sa2-1979}, which was further improved by using a depth-independent effective complex matrix of Raman tensor to approximate the depth-dependent polarization \cite{Kranert-prl-2016,kranert-srep-2016}. For opaque anisotropic crystals (OAC), not only birefringence effect but also dichroism effect from anisotropic absorption is responsible for polarized Raman scattering. With the boomed emergence of in-plane ALMs \cite{qiao-nc-2014,wang-jacs-2017,Zhou2019JOS,li-infor-2019}, $e.g.$, black phosphorus (BP), the investigations of angle-resolved polarized Raman (ARPR) intensity in OAC have received great attention \cite{huang-acsnano-2016,Ribeiro-ACSNano-2015,Kim-nanoscale-2015,Mao-small-2016,Ling-Nanolett-2016,zhang-acsnano-2017,Choi2020NH}.

In analogy to Raman scattering in isotropic crystal under a backscattering configuration, the ARPR intensity in OAC is usually estimated by its Raman tensor and polarization of incident laser and scattered signal outside the crystals by introducing a fitted complex Raman tensor due to dichroism \cite{strach-prb-1998} or a fitted birefringence-induced phase delay between electric field components \cite{Alonso-prb-2005}. These two semi-quantitative approaches \cite{strach-prb-1998,Alonso-prb-2005} are commonly utilized to reproduce the ARPR intensity dependent on the laser wavelength and thickness of ALM flakes in backscattering geometry at normal laser incidence \cite{huang-acsnano-2016,Ribeiro-ACSNano-2015,Kim-nanoscale-2015,Mao-small-2016,Ling-Nanolett-2016,zhang-acsnano-2017,Choi2020NH}. Although the two approaches are, respectively, associated with birefringence and dichroism, the mathematical expressions for ARPR intensity of a Raman mode are the same at normal laser incidence \cite{Ribeiro-ACSNano-2015,Mao-small-2016}, leading to ambiguous origins for the corresponding anomalous ARPR intensity of ALMs. For simplicity, we denoted these two approaches as phase-difference-based (PDB) model. Furthermore, these approaches can not be applicable to the case of ARPR intensity at oblique laser incidence because of the complex depth-dependent polarization and intensity of incident laser and scattered signal inside ALMs, and additional angle-dependent reflection and refraction at the interface between ALM flakes and air. This is the reason why the birefringence and dichroism effects on Raman intensity is $''$predictably catastrophic$''$ \cite{DAWSON-sa-1972}. Fundamentally, only real Raman tensor is generally involved if no magnetic perturbation occurs \cite{Long-book-1977}. Indeed, only real Raman tensor is considered to deduce a formalism for calculating the Raman scattering intensity dependent on the polarization configuration for optically anisotropic crystals \cite{Kranert-prl-2016,kranert-srep-2016}. Thus, this leaves an open question whether it is possible to reproduce ARPR intensity of OAC by only the real Raman tensor, especially for emergent ALMs. Though more than 50 years has been past since the unusual polarized Raman intensity in anisotropic crystals was noticed \cite{porto-pr-1966}, there is still no general method to quantitatively calculate the ARPR intensity with both birefringence and linear dichroism considered.

Here, by taking BP as an example, we provided a model to understand the ARPR intensity at normal and oblique laser incidences on in-plane ALMs. In this model, birefringence and linear dichroism are considered by different complex refractive indexes along three principle axes, which is directly measured from incident-angle resolved reflectivity. This results in depth-dependent polarization and intensity of laser at the scattering site and of the corresponding Raman signal inside the crystal. The ARPR intensity at normal and oblique laser incidences can be understood by the same set of real Raman tensors for certain laser excitation without any fitting parameter.

\section{Materials and Methods}
\subsection{Sample preparations}
The BP crystals were synthesized by using a modified chemical vapor transport method with raw materials including amorphous red phosphorus, tin, and iodine \cite{KOPF-jcg-2014}. The as-prepared crystals were washed with acetone and ethanol to remove the surface adsorbed iodine. The typical optical images of the synthesized bulk BP in Fig. S1 shows a smooth surface of the sample with long straight edges, which implies the high quality of the bulk BP with negligible oxidative degradation. This promises the possibility to measure its complex refractive index by incident-angle resolved reflectivity according to the Fresnel$'$s law, which would be discussed later.

\subsection{Raman spectroscopy and incident-angle dependent intensity reflectivity measurements}

The Raman spectra were measured in backscattering geometry using a Jobin-Yvon HR-Evolution micro-Raman system equipped with an edge filter and a charge-coupled device detector. The excitation wavelengths are 532 nm from a solid state laser and 488 nm from a Ar$^+$ laser. For ARPR measurements at normal laser incidence, a 100$\times$ microscope objective was used for focusing the incident light and collecting the scattered Raman signal. A polarizer was placed in the incident path of the Raman instrument, and the analyzer with polarization parallel to that of the polarizer was allocated before the spectrometer. A half-wave plate is inserted in the common optical path of incident and scattered light to simultaneously vary their polarization directions. By rotating the fast axis of the half-wave plate with an angle of $\theta$/2, the polarization of incident and scattered light is rotated by $\theta$ relative to the $x$ axis of BP. The laser power is less than 150 $\mu$W to avoid heating the BP samples.

For the case of oblique laser incidence, the samples were placed vertically on a rotating platform, so that the incident angle is tunable. The laser beam propagates through a half-wave plate, and is then focused onto the sample by a convex lens with numerical aperture (NA) of 0.04, leading to an angle resolution of 4.3$^{\circ}$ for the incident-angle resolved Raman scattering. The scattered Raman signal travels back and its polarization is also rotated by the half-wave plate before being selected by an analyzer with polarization parallel to that of the polarizer. By rotating the half-wave plate, the polarization of laser and collected Raman signals can be simultaneously vertical (in-plane configuration) or horizontal (out-of-plane configuration), which would be detailed demonstrated below. Meanwhile, the reflected light is collected by a single channel detector (SCD). The incident-angle dependent intensity reflectivity of $s$- ($R_s$) and $p$-polarization ($R_p$) components can be obtained by the relative intensity of reflected light to the incident light in in-plane and out-of-plane configuration, respectively, in which the $s$($p$)-polarization is perpendicular (parallel) to the incident plane. The intensity of the laser was less than 2.0 mW to avoid heating the BP samples.

\section{Results and Discussion}
\subsection{Raman tensors, complex refractive indexes and ARPR intensity of BP at normal incidence by the BLD model}

\begin{figure*}[!htb]
\centerline{\includegraphics[width=180mm,clip]{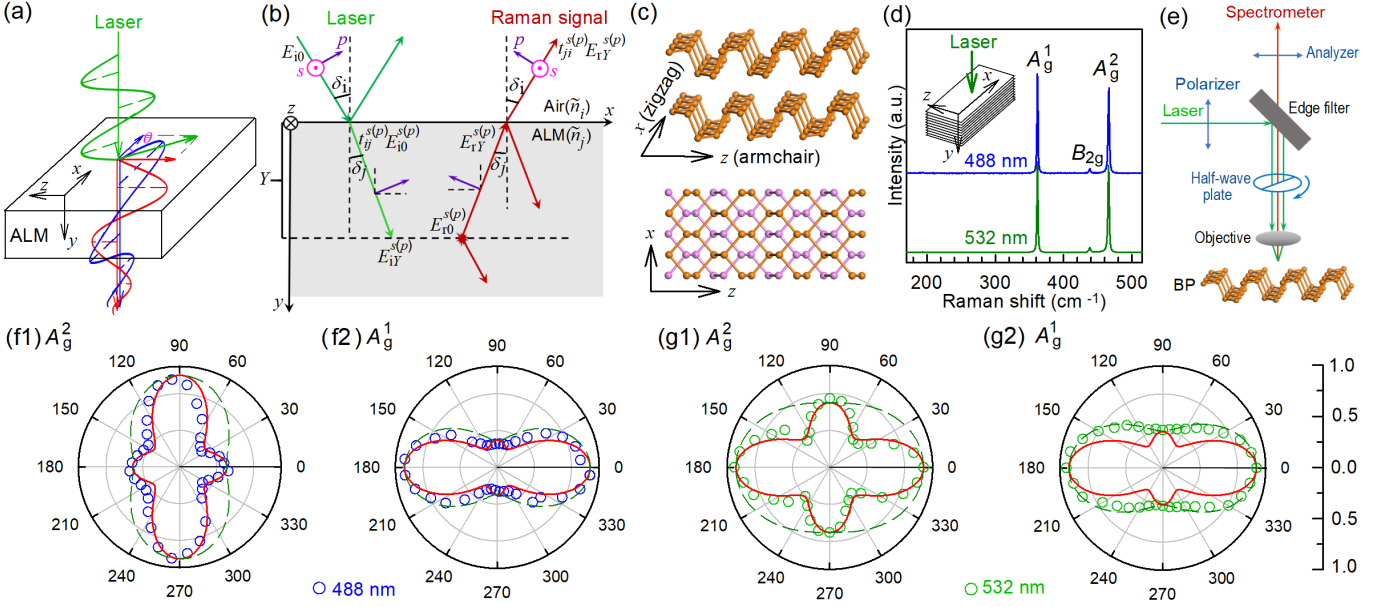}}
\caption{(Color online) ARPR spectra of BP at normal incidence to the basal plane. (a) Schematic illustration of the birefringence and linear dichroism , (b) propagation path of laser (green) and scattered Raman signal (dark red) in ALM flakes. (c) Schematic diagrams of BP from the side and top view. (d) Raman spectra of BP with lasers of 488 nm (blue) and 532 nm (green) normal incidence to its basal plane. (e) The setup for ARPR measurements. ARPR intensities of the $A_{\rm g}^2$ and $A_{\rm g}^1$ modes excited by (f) 488 nm and (g) 532 nm lasers under the parallel polarization configuration. The solid and dashed lines are the calculated results by the BLD model and the approach without considering birefringence and linear dichroism, respectively.}\label{Fig-1}
\end{figure*}

For a general Raman scattering, the intensity is associated with a real Raman tensor based on Raman selection rule \cite{Loudon-book-1964},
\begin{equation}
I \propto \sum\left|\textit{\textbf{e}}_{\rm r}\cdot {\rm \textbf{R}}\cdot \textit{\textbf{e}}_{\rm i}\right|^2,
\label{eq1}
\end{equation}
\noindent where \textbf{R} is the Raman tensors of a phonon mode, \textit{$\textbf{e}_{\rm i}$} and \textit{$\textbf{e}_{\rm r}$} represent polarization vectors of incident laser and Raman signal at the scattering site inside crystal, respectively. For isotropic crystals, \textit{$\textbf{e}_{\rm i}$} and \textit{$\textbf{e}_{\rm r}$} can be approximatively replaced by the polarization vectors of incident laser and collected Raman signal, respectively, to fit the experimental results. However, the case within ALMs would be much complicated, because birefringence and linear dichroism would, respectively, lead to varied phase velocities and penetration depths along different directions, as illustrated in Fig.\ref{Fig-1}(a). Thus, the polarization and intensity of incident laser and Raman signal are depth-dependent inside ALMs, especially for oblique laser incidence with an incident angle of $\delta_{\rm i}$, as indicated in Fig.\ref{Fig-1}(b). The electric field vector of the incident laser can be classified into $s$- ($E^{s}_{\rm i0}$) and $p$-polarization ($E^{p}_{\rm i0}$) components. The electric field of incident laser that passes through the interface is $t^{s(p)}_{ij}E^{s(p)}_{\rm i0}$, where $t^{s(p)}_{ij}$ is the amplitude transmission coefficient for $s$($p$)-polarization component from medium $i$ (air) to $j$ (ALMs). It would experience depth-dependent phase delay and intensity; thus, at site $Y$ within ALMs, it becomes $E^{s(p)}_{{\rm i}Y}$. A similar case occurs for the Raman signal ($E^{s(p)}_{\rm r0}$) at site $Y$, and the corresponding transmission coefficient is $t^{s(p)}_{ji}$ due to the opposite direction. For simplicity, we denoted the approach to understand the ARPR intensity of ALMs by both the birefringence and linear dichroism effects as birefringence-linear-dichroism (BLD) model.

To give an insight into the polarization selection rule for ALMs at normal and oblique laser incidences by BLD model, we take BP as a prototype, because its in-plane zigzag and armchair directions exhibit different complex refractive indexes, leading to obvious birefringence and linear dichroism \cite{Asahina-JPC-1984,qiao-nc-2014}. Following the most common definition, $x$ and $z$ axes are along zigzag and armchair directions of BP, respectively, leaving the $y$ axis perpendicular to the basal plane \cite{Sugai-SSC-1985}, as shown in Fig.\ref{Fig-1}(c). Six normal phonon modes ($2A_{\rm g}+B_{\rm 1g}+B_{\rm 2g}+2B_{\rm 3g}$) at $\Gamma$ point are Raman active. The Raman tensor \textbf{R} with elements $R_{uv}$ ($u,v=x,y,z$) for the two $A_{\rm g}$ modes are

\begin{equation}
\begin{aligned}
&R(A_{\rm g})={
\left( \begin{array}{ccc}
a & 0 & 0\\
0 & b & 0\\
0 & 0 & c
\end{array}
\right )}.
\end{aligned}
\label{eq2}
\end{equation}\\
\noindent Our synthesized BP samples exhibit perfect zigzag and armchair (side) planes, whose Raman spectra excited by 488 nm and 532 nm lasers are plotted in Fig.\ref{Fig-1}(d). The three peaks at 362, 436, and 466 cm$^{-1}$ can be assigned to the $A_{\rm g}^1$, $B_{\rm 2g}$ and $A_{\rm g}^2$ modes \cite{Kaneta-ssc-1982}, respectively.

To reproduce the ARPR intensity at normal and oblique laser incidences, one fundamental question is how to obtain the Raman tensor of the two $A_{\rm g}$ modes. We started with the ARPR measurements at normal laser incidence ($\delta_{\rm i}=0^\circ$) on the basal plane under a parallel polarization configuration, as illustrated in Fig.\ref{Fig-1}(e). The relative angle $\theta$ between laser polarization vector and the $x$ axis was controlled by a half-wave plate. At $\theta$ = 0$^{\circ}$ (180$^{\circ}$) and 90$^{\circ}$(270$^{\circ}$), the laser polarization is along the $x$ and $z$ axes, respectively. The measured ARPR intensity of the two modes under the backscattering configuration are summarized in Fig.\ref{Fig-1}(f-g). The ARPR intensity of the $A_{\rm g}^2$ mode exhibits wavelength-dependent profiles, showing maxima along $z$ and $x$ axes for 488- and 532-nm lasers, respectively. However, the profile of the $A_{\rm g}^1$ mode is similar for the two lasers. This can not be understood by the general Raman scattering for isotropic crystals, implying that the birefringence and linear dichroism should be taken into account for the unusual ARPR intensity of BP.

\begin{figure*}[!htb]
\centerline{\includegraphics[width=180mm,clip]{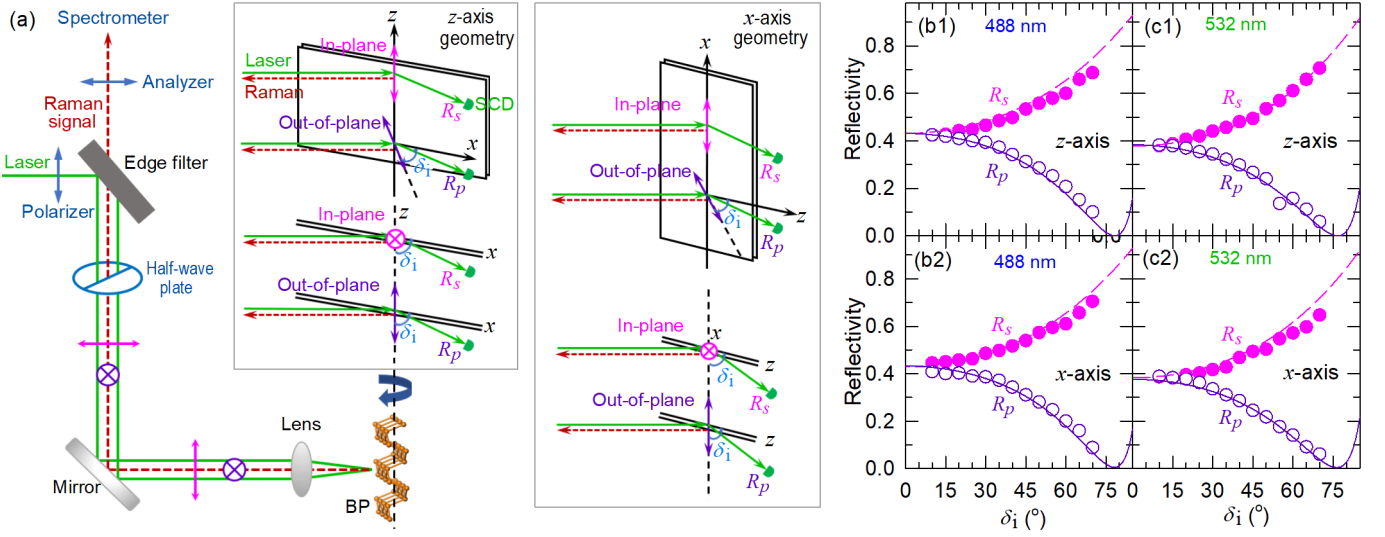}}
\caption{(Color online) Angle-resolved reflectivity at oblique laser incidence. (a) Schematic setup to measure angle-resolved polarized reflectivity at oblique laser incidence. SCD, single channel detector. The insets show schematic diagrams of in-plane and out-of-plane configurations for the $z$-axis (left) and $x$-axis (right) rotation geometries in a side and top view. (b,c) The experimental reflectivity (filled and open circles) and fitted curves (dashed and solid lines) of $s$- ($R_s$) and $p$-polarization ($R_p$) laser, respectively, under in-plane (pink) and out-of-plane (purple) configurations. (b1,c1) and (b2,c2) shows data in $z$- and $x$-axis rotation geometries, respectively.}\label{Fig-2}
\end{figure*}

Birefringence and linear dichroism in BP result from different complex refractive indexes, $\tilde{n}_x$, $\tilde{n}_z$ and $\tilde{n}_y$, respectively, along zigzag, armchair and out-of-plane axes. Accurate $\tilde{n}_x$, $\tilde{n}_y$ and $\tilde{n}_z$ are crucial to understand the ARPR intensity influenced by birefringence and linear dichroism. The perfect basal plane of the BP sample (Fig. S1) make it possible to measure its $\tilde{n}$ by $\delta_{\rm i}$-resolved reflectivity according to the Fresnel$'$s law (Supplementary Section 1). As illustrated in Fig.\ref{Fig-2}(a), the sample is placed vertically on a rotating platform to rotate along its $z$ axis by $\delta_{\rm i}$ relative to the $y$ axis, denoted as $z$-axis rotation geometry. The laser beam propagates through a half-wave plate, and is then focused onto sample surface by a convex lens with a NA of 0.04. With this setup, the ARPR signal can be collected by the same convex lens and probed by spectrometer, while the reflected laser can be directly detected by a SCD. By rotating the half-wave plate, the polarization of oblique incident laser can be either vertical (pink arrows) or horizontal (purple arrows), corresponding to the $s$- and $p$-polarizations, respectively. And the corresponding configurations can be respectively defined as in-plane and out-of-plane configurations in terms of the BP basal plane. Similarly, the sample can rotate $\delta_{\rm i}$ along its $x$ axis, $i.e.,$ $x$-axis rotation geometry (right panel of insets in Fig.\ref{Fig-2}(a)). In the case of out-of-plane configuration, the $p$ polarization presents an angle of $\delta_{\rm i}$ relative to the $x$ ($z$) axis in $z$($x$)-axis geometry. $\delta_{\rm i}$-dependent $R_s$ and $R_p$ can be obtained by the relative intensity between the reflected and incident laser, as shown in Fig.\ref{Fig-2}(b,c) for 488 and 532 nm lasers under the two configurations in both $x$- and $z$-axis rotation geometries. All the experimental results can be well fitted with the Fresnel$'$s law and the fitted $\tilde{n}_x$, $\tilde{n}_y$ and $\tilde{n}_z$ are listed in Table \ref{tb1}. The much smaller imaginary part along $x$ axis than that along $z$ axis indicates its much lower absorption efficiency, which is similar to the previous results \cite{Asahina-JPC-1984,yuan-nn-2015,zhang-nc-2017,qiao-nc-2014}.

\begin{table}[htbp]
\centering\caption{Complex refractive indexes $\tilde{n}$ along three principle axes of bulk BP at 488 nm and 532 nm.}
\begin{tabular}{cccc}
\hline
Wavelength & $\tilde{n}_x$ & $\tilde{n}_y$ & $\tilde{n}_z$\\
\hline
488 nm & 4.82+0.067i & 4.62+0.73i & 4.72+0.71i\\
532 nm & 4.25+0.054i & 4.06+0.32i & 4.10+0.55i\\
\hline
\end{tabular}
\label{tb1}
\end{table}

With the obtained $\tilde{n}$ from experiments as discussed above, we can go back to analyze the ARPR data in Fig.\ref{Fig-1}(f-g). In the case of normal laser incidence, the $s$- and $p$-polarization components are respectively along $z$ and $x$ axes in a view of the zigzag side plane. The linear polarized incident laser at BP surface can be simply decomposed into two components along its $x$ and $z$ axes, $i.e.$, $E^x_{\rm i0}=E_{\rm i0}{\rm cos}\theta$ and $E^z_{\rm i0}=E_{\rm i0}{\rm sin}\theta$ \cite{Jones-JOSA-1941,Liu-cpb-2017}, where $E_{\rm i0}$ is the electric field of the incident laser. Notably, $t^{s(p)}_{ij}$ is neglected due to the small difference between $s$- and $p$-polarizations at normal incidence. The two components propagate with different phase velocities and penetration depths within BP. With a propagation distance of $Y$, they become,

\begin{equation}
\begin{split}
& E^x_{{\rm i}Y}=E_{\rm i0}{\rm e}^{2\uppi i\tilde{n}_xY/\lambda} {\rm cos}\theta=E_{\rm i0} e^x_{{\rm i}Y},\\
& E^z_{{\rm i}Y}=E_{\rm i0}{\rm e}^{2{\rm \uppi} i\tilde{n}_zY/\lambda}{\rm sin}\theta=E_{\rm i0} e^z_{{\rm i}Y},
\end{split}
\label{eq4}
\end{equation}

\noindent where $\lambda$ is the excitation wavelength. The electric field components of Raman signal ($E^u_{\rm r0}$, $u=x,y,z$) can be given by $E^v_{{\rm i}Y}$ ($v=x,y,z$) and the Raman tensor, $i.e., E^u_{\rm r0}=\sum_{v=x,y,z}R_{uv}E^v_{{\rm i}Y}$. Each component propagates with respective phase velocity and penetration depth; thus, the electric field components of the collected Raman signal should be $E^u_{{\rm r}Y}=E^u_{\rm r0}e^u_{{\rm r}Y}$($u=x,y,z$), where $e^u_{{\rm r}Y}=e^u_{{\rm i}Y}$ for parallel polarization configuration (Supplementary Section 2). Therefore, the intensity of scattered Raman signal is obtained by summing the squares of electric field components integrating over the penetration depth,\\

\begin{equation}
 I\propto\sum_{u=x,y,z}\left|\int_{0}^{\infty}E^u_{{\rm r}Y}dY\right|^2.
\label{eq5}
\end{equation}

\noindent Because of the small phonon energy, the differences in wavelength and complex refractive index between incident and scattered light are neglected.

\begin{table}[!htbp]
\centering
\caption{Raman tensor elements $R_{uv}$ of the $A_{\rm g}^2$ and $A_{\rm g}^1$ modes of BP at 488 nm and 532 nm.}
\begin{tabular}{ccccccccc}
\hline
\multicolumn{2}{c}{\multirow{2}{*}{Wavelength}}&
\multicolumn{3}{c}{$A_{\rm g}^2$} &\,&
\multicolumn{3}{c}{$A_{\rm g}^1$} \\
\cline{3-5} \cline{7-9}
\multicolumn{2}{c}{}&$a$&$b$&$c$&\,&$a$&$b$&$c$\\
\hline
\multicolumn{2}{c}{488 nm}&1.00&0.81&2.70&\,&1.00&0.53&1.05\\
\multicolumn{2}{c}{532 nm}&1.00&2.40&0.56&\,&1.00&2.40&0.42\\
\hline
\end{tabular}
\label{tb2}
\end{table}

Because there is no birefringence effect along $z$ and $x$ axes, based on Eq.\ref{eq5}, we can obtain $I(\theta=0^\circ(180^\circ))\propto a^2$ and $I(\theta=90^\circ(270^\circ))\propto c^2$. Thus, with the known $\tilde{n}_x$ and $\tilde{n}_z$ in Table \ref{tb1}, the real Raman tensor elements of $a$ and $c$ of the two Raman modes can be deduced from the measured ARPR intensity at 0$^\circ$(180$^\circ$) relative to that at 90$^\circ$(270$^\circ$) at normal laser incidence on BP in Fig.\ref{Fig-1}(f-g). The corresponding elements for 488-nm and 532-nm lasers are summarized in Table \ref{tb2}. Based on these  Raman tensors, one can calculate the whole ARPR intensity in the range of 0$\sim$360$^\circ$ by Eq.\ref{eq5}, as shown in Fig.\ref{Fig-1}(f-g) by solid lines, which are in line with the experimental ARPR intensity of the $A_{\rm g}^2$ mode for both 488-nm and 532-nm lasers. The solid line is not perfectly fitted for the $A_{\rm g}^1$ mode excited by 532 nm laser, which may be ascribe to the complex electron-photon and electron-phonon coupling related to the $A_{\rm g}^1$ mode in the resonance region. By fitting the ARPR intensity with Eq. \ref{eq1} without considering the birefringence and linear dichroism effects, the corresponding results are depicted by the dashed lines in Figs.\ref{Fig-1}(f-g). The significant difference of the $A_{\rm g}^2$ mode between solid and dashed lines in Figs.\ref{Fig-1}(f1)and (g1) indicates the necessity to include the birefringence and linear dichroism to explain the experimental ARPR data.

\begin{figure}
\centerline{\includegraphics[width=90mm,clip]{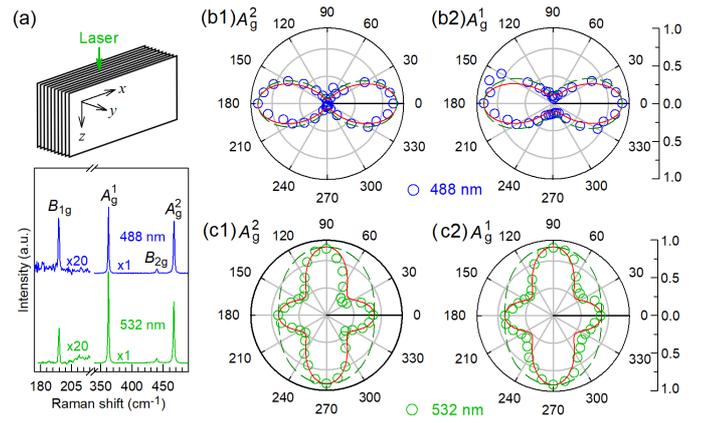}}
\caption{(Color online) ARPR spectra of BP at normal incidence on the side plane. (a) Raman spectra from the zigzag side plane of BP excited by 488 and 532 nm. ARPR intensities of the $A_{\rm g}^2$ and $A_{\rm g}^1$ modes excited by (b1,b2) 488 and (c1,c2) 532 nm  lasers. The solid and dahsed lines depict the calculated ARPR results by the BLD model and the approach without considering birefringence and linear dichroism, respectively.}\label{Fig-3}
\end{figure}

To get the Raman tensor element $b$ of the $A_{\rm g}$ modes, ARPR spectra of BP at the zigzag side plane are measured and typical spectra are shown in Fig.\ref{Fig-3}(a), where $B_{\rm 1g}$ mode becomes Raman active and appears. The ARPR intensities of the $A_{\rm g}^2$ and $A_{\rm g}^1$ modes (circles) were also measured at the zigzag side plane with parallel polarization configuration in Fig.\ref{Fig-1}(e), as shown in Fig. \ref{Fig-3}(b-c). The relative value between elements $a$ and $b$ for the two $A_{\rm g}$ modes can also be deduced from the measured ARPR intensity at 0$^\circ$(180$^\circ$) relative to that at 90$^\circ$(270$^\circ$) at normal laser incidence. The obtained Raman tensor elements are also included in Table \ref{tb2} for both 488 and 532 nm lasers. All the data in Fig. \ref{Fig-3}(b-c) can be reproduced by Eq.\ref{eq5} (solid lines) based on the obtained real Raman tensors at the two laser wavelengths. In contrast, the two $A_{\rm g}$ modes excited by 532 nm on the side plane in Figs.\ref{Fig-3}(c1,c2) are not in agreement with the calculation based on Eq.\ref{eq1} (dashed lines) without taking birefringence and linear dichroism into account. The above results (Figs.\ref{Fig-1}(f-g) and \ref{Fig-3}(b-c)) suggest that, the real Raman tensors of the two $A_{\rm g}$ modes of BP can be deduced by the relative Raman intensity along three principle axes at normal laser incidence on the basal and side planes. They can be used to reproduce the ARPR intensity at normal laser incidence on the basal and side planes of BP if one takes the depth-dependent polarization and intensity into account, which result from the effects of birefringence and linear dichroism in BP. This method to determine real Raman tensor in BP is applicable to all ALMs once the complex refractive indexes of the three principle axes are known, which provides a valuable method for the prediction of ARPR intensity and further the characterization of crystallographic orientation.

\subsection{ARPR intensity at oblique laser incidence by the BLD model}

\begin{figure*}[!htb]
\centerline{\includegraphics[width=180mm,clip]{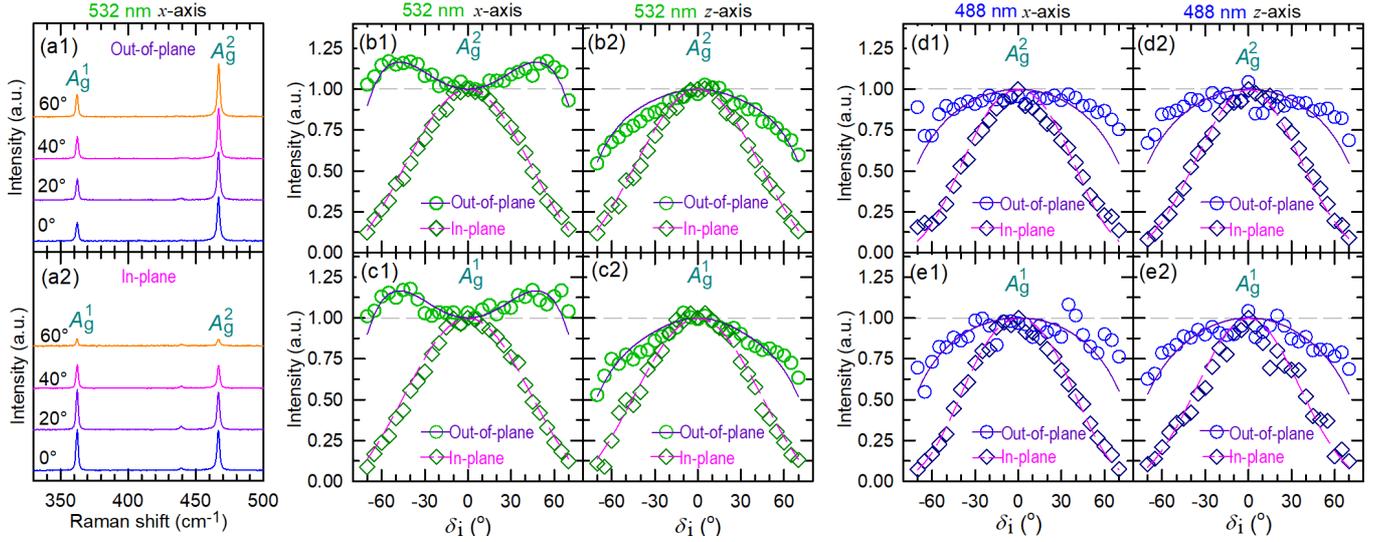}}
\caption{(Color online) ARPR spectra of BP at oblique laser incidence. (a) Typical Raman spectra at several $\delta_{\rm i}$ under (a1) out-of-plane and (a2) in-plane configurations excited by 532 nm in $x$-axis rotation geometry. The experimental (open diamonds and circles) and calculated (dashed and solid lines) $\delta_{\rm i}$-dependent ARPR intensity of the $A_{\rm g}^2$ and $A_{\rm g}^1$ modes excited by 532 nm laser under in-plane and out-of-plane configurations in $x$- (b1,c1) and $z$-axis (b2,c2) rotation geometries, respectively. The corresponding results excited by 488 nm laser are shown in (d,e).}\label{Fig-4}
\end{figure*}

The ARPR spectra at oblique laser incidence are much complicated than the case of normal incidence. For the oblique incidence, the $\delta_{\rm i}$-dependent ARPR intensity can be measured using the setup in $z$- and $x$-axis rotation geometries schematized in Fig.\ref{Fig-2}(a). The typical ARPR spectra under 532 nm excitation in the latter case are plotted in Fig.\ref{Fig-4}(a). The corresponding ARPR intensity of the $A^2_{\rm g}$ and $A^1_{\rm g}$ modes in both $x$- and $z$-axis rotation geometries are summarized in Fig.\ref{Fig-4}(b) and (c), respectively. For the in-plane configuration, the Raman intensity for the $A^2_{\rm g}$ and $A^1_{\rm g}$ modes (open diamonds) decreases monotonically with $\delta_{\rm i}$ increasing, regardless of the rotation geometries. In contrast, for out-of-plane configuration, the $\delta_{\rm i}$-dependent ARPR intensity of the two modes (open circles) is sensitive to the rotation axis, and that for the $x$-axis geometry exhibits a maximum at $\delta_{\rm i}\sim50^{\circ}$, as shown in Fig.\ref{Fig-4}(b1,c1).

To understand the unusual ARPR intensity at oblique laser incidence, not only the depth-dependent polarization and intensity within the sample but also the $\delta_{\rm i}$-dependent $t^{s(p)}_{ij}$ ($t^{s(p)}_{ji}$) should be considered due to the $\delta_{\rm i}$-dependent reflectivity as addressed in Fig.\ref{Fig-2}, which can be calculated by the Fresnel$'$s law (Supplementary Section 3 and 4). For in-plane configuration with $x$-axis rotation geometry, $\tilde{n}_s$=$\tilde{n}_x$. Because of small NA of the used convex len, the optical path of excitation beam and Raman signal is almost the same. Thus, the electric field components of incident light at $Y$ and the corresponding scattered Raman signal collected outside BP are

\begin{equation}
E_{{\rm i}Y}^s = t_{ij}^s E_{\rm i0} {\rm e}^{2 {\rm \uppi} i\tilde{n}_s Y / \lambda},
E_{{\rm r}Y}^s = t_{ji}^s E^s_{\rm r0}{\rm e}^{2 {\rm \uppi} i\tilde{n}_s Y / \lambda},
\label{eq6}
\end{equation}

\noindent respectively. According to Eqs. \ref{eq5} and \ref{eq6}, the $\delta_{\rm i}$-dependent Raman intensity for in-plane configuration excited by 532 nm laser can be calculated with the known Raman tensors and complex refractive indexes, as depicted by dashed lines in Fig.\ref{Fig-4}(b,c), showing perfect agreement with the experimental results.

In the case of out-of-plane configuration, the ARPR intensity can also be calculated similar to the case of in-plane configuration while the incident laser and scattered signal provides electric field components along $y$ and $z$ directions,
\begin{equation}
\begin{split}
E_{{\rm i}Y}^{yp}(E_{{\rm i}Y}^{zp})= t_{ij}^p E_{\rm i0}{\rm e}^{2\uppi i\tilde{n}_pY/\lambda}{\rm sin}\delta_j({\rm cos}\delta_j),\\
E_{{\rm r}Y}^{yp}(E_{{\rm r}Y}^{zp})= t_{ji}^p E_{\rm r0}^{yp}(E_{\rm r0}^{zp}) {\rm e}^{2 \uppi i\tilde{n}_p Y / \lambda} {\rm sin}\delta_j({\rm cos}\delta_j),
\end{split}
\label{eq7}
\end{equation}
\noindent where $\delta_j$ is the refraction angle related to $\delta_{\rm i}$, $\tilde{n}^2_p = \tilde{n}_{\rm para}^2 \tilde{n}_{\rm perp}^2/(\tilde{n}_{\rm para}^2 {\rm sin}^2\delta_j + \tilde{n}_{\rm perp}^2 {\rm cos}^2\delta_j)$ is the complex refractive index of $p$-polarization in BP, with $\tilde{n}_{\rm perp}$ and $\tilde{n}_{\rm para}$ representing the complex refractive indexes perpendicular ($\tilde{n}_y$) and parallel ($\tilde{n}_z$ or $\tilde{n}_x$) to the basal plane, respectively. The calculated $\delta_{\rm i}$-dependent Raman intensities for out-of-plane configuration excited by 532 nm laser depicted by solid lines in Fig.\ref{Fig-4}(b,c) are in line with the experimental results for the two rotation geometries, although the polarized behaviors are quite different between them. The corresponding experimental and calculated results excited by 488 nm for the two rotation geometries shown in Fig.\ref{Fig-4}(d,e) are also in agreement with each other, where the experimental ARPR intensity in $x$-axis geometry is significantly different from that excited by 532 nm laser.

\subsection{Discussion}

The evidently different ARPR intensity in $x$- and $z$-axis rotation geometries implies the great anisotropy in BP. The agreements between the experimental data and the calculated results based on the BLD model further confirms the influence of both birefringence and linear dichroism on the ARPR intensity. It should be noted that the calculated ARPR intensity at oblique laser incidence for both in-plane and out-of-plane configurations by the BLD model is only related to $\tilde{n}$ and real $\textbf{R}$ from Tables \ref{tb1} and \ref{tb2}, respectively, without any fitting parameters in the calculation. Thus, the ARPR intensity for normal and oblique incidence to ALM flakes can be quantitatively described $via$ real $\textbf{R}$ by considering the corresponding depth-dependent polarization and intensity in ALMs induced by both birefringence and linear dichroism effects based on the BLD model.

Nowadays, the PDB model is commonly utilized to fit the experimental ARPR intensity in ALMs, especially for ultrathin ALM flakes deposited on multilayered substrates \cite{Ribeiro-ACSNano-2015,Kim-nanoscale-2015,Mao-small-2016,huang-acsnano-2016,Ling-Nanolett-2016,zhang-acsnano-2017,Choi2020NH}. However, the PDB model can be only applied to the ARPR intensity at normal incidence by considering the polarization of incident laser and scattered signal outside the crystals, similar to the case in isotropic materials. Accordingly, the complex Raman tensors \cite{Ribeiro-ACSNano-2015} in bulk BP can be obtained by fitting the experimental ARPR intensity at normal incidence, as compared with the real Raman tensors from the BLD model in Table S2. In the PDB model, $\phi_{ca}$($\phi_{ba}$) is the phase difference between the two components along $x$ and $z$ ($y$) axis \cite{Ribeiro-ACSNano-2015}. The absolute value of Raman tensor element $a$ is fixed at 1.00. When the ARPR intensity shows one maximum along a principle axis, $e.g.,$ $A_{\rm g}^1$ and $A_{\rm g}^2$ modes excited by 488 nm laser in Fig. \ref{Fig-3}(b), the ratio between $a$ and corresponding $b$ (or $c$) would be large while the corresponding phase difference is close to zero. In this model, the birefringence or linear dichroism effects are simplified by the effective formalism of Raman tensor elements and phase difference. At a given laser wavelength, Raman tensor elements and phase difference should depend on the thickness of ALM flakes and dielectric layers of substrate. Thus, at normal laser incidence, although the complex Raman tensors of a specific ALM flake on a multilayered substrate can be obtained by fitting the measured ARPR intensity data, it is difficult to foresee the ARPR intensity of the ALM flake by the obtained complex Raman tensors once the thickness of ALM flake or dielectric layers of substrate is changed.

The PDB model is also difficult for predicting the ARPR intensity of ALMs at oblique laser incidence by the complex Raman tensors obtained at normal incidence. This can be ascribed to the varied $\phi_{ca}$ and $\phi_{ba}$ of the $A_{\rm g}^1$ and $A_{\rm g}^2$ modes at normal and oblique incidences, which should be dependent on in-plane/out-of-plane configuration and $\delta_{\rm i}$. If we rigidly apply these fitted complex Raman tensors at normal incidence to the case of oblique laser incidence, the ARPR intensity can be further calculated by Eq. 1, as depicted in Fig. S2. Obviously, the calculated ARPR intensities rigidly based on the PDB model show discrepancy with the experimental results, especially for the out-of-plane configuration. It should be noted that the $\delta_{\rm i}$-dependent reflectivity (Fig.\ref{Fig-2}(b,c)) at the BP/air interface is also considered in the above calculation. Otherwise, the ARPR intensity would remain constant for in-plane configuration since only the electric field component along $x$ or $z$ axis is involved, which obviously disagrees with the experimental results.

In contrast, based on the BLD model, the agreement between experimental results and predicted ARPR intensity at normal and oblique laser incidences indicate that it is possible to quantitatively reproduce the ARPR intensity of all ALMs for a given excitation wavelength under any scattering and polarization configurations. This can be extended to any OAC even to a two-dimensional case, $e.g.$, ultrathin ALM flakes once their real $\textbf{R}$ and $\tilde{n}$ are known. It is worth mentioning that the interference effects in the multilayered structures containing air, ALM flakes and substrate for the incident laser and out-going Raman signals should be considered in the two-dimensional case. The well-known anomalous ARPR intensity in ALM flakes \cite{Ribeiro-ACSNano-2015,Kim-nanoscale-2015,Mao-small-2016,huang-acsnano-2016,Ling-Nanolett-2016,zhang-acsnano-2017,Choi2020NH} sensitive to the laser wavelength, thickness of ALM flakes and dielectric layers of the substrate would be expected to be understood by the BLD model if one considers the depth-dependent polarization and intensity of incident laser and scattered Raman signal induced by both birefringence and linear dichroism effects within ALM flakes and the interference effects in the multilayered structures. However, it is beyond the scope of this work and will be discussed in the future work.

\section{Conclusion}

In conclusion, based on the proposed BLD model, we quantitatively understand the ARPR intensity in ALMs $via$ real Raman tensor by considering depth-dependent polarization and intensity of incident laser and scattered Raman signal induced by birefringence and linear dichroism effects. With the real $\textbf{R}$ obtained from the relative Raman intensity along its principle axes at normal laser incidence and $\tilde{n}$ of three principle axes from incident-angle resolved reflectivity, this method have successfully reproduced the angle-resolved Raman intensity in BP at both normal and oblique laser incidences, where the transmission coefficient is considered for the latter case. This work can be extended to other materials with strong birefringence and linear dichroism, shedding light on the quantitative analysis for polarized Raman scattering in anisotropic materials.

\section{Conflict of interest}

The authors declare that they have no conflict of interest.

\section{Acknowledgments}
We acknowledge support from the National Key Research and Development Program of China (Grant No. 2016YFA0301204), the National Natural Science Foundation of China (Grant Nos. 11874350 and 51702352), the CAS Key Research Program of Frontier Sciences (Grant NO. ZDBS-LY-SLH004) and China Postdoctoral Science Foundation (Grant No. 2019TQ0317). J. Wang acknowledges support from Youth Innovation Promotion Association Chinese Academy of Sciences (2020354).

\section{Author Contributions}

Ping-Heng Tan conceived the idea, directed and supervised the project. Yu-Chen Leng and Ping-Heng Tan designed the experiments. Yu-Chen Leng performed experiments. Jiahong Wang, Binlu Yu and Xue-Feng Yu prepared the samples. Miao-Ling Lin, Yu-Chen Leng and Ping-Heng Tan analyzed the data with inputs from Da Meng, Xin Cong, Xiao-Li Li and Xue-Lu Liu; Yu-Chen Leng, Miao-Ling Lin and Ping-Heng Tan developed the theoretical model. Miao-Ling Lin, Yu-Chen Leng, and Ping-Heng Tan wrote the manuscript with input from all authors.

\section{Appendix A. Supplementary materials}

Supplementary data to this article can be found online at https://doi.org/xxx.

\bibliographystyle{csb}
\bibliography{BP-ref}

\end{document}


\begin{frontmatter}



\title{Supplementary Materials for\\
Understanding angle-resolved polarized Raman scattering from black phosphorus at normal and oblique laser incidences}

\author[a]{Miao-Ling Lin\fnref{1}}
\author[a,b]{Yu-Chen Leng\fnref{1}}
\author[a,b]{Da Meng}
\author[a,b]{Xin Cong}
\author[c]{Jiahong Wang}
\author[d]{Xiao-Li Li}
\author[c]{Binlu Yu}
\author[a]{Xue-Lu Liu}
\author[c]{Xue-Feng Yu}
\author[a,b,e]{Ping-Heng Tan\corref{cor1}}
\ead{phtan@semi.ac.cn}
\cortext[cor1]{Corresponding author.}
\address[a]{State Key Laboratory of Superlattices and Microstructures, Institute of Semiconductors, Chinese Academy of Sciences, Beijing 100083, China}
\address[b]{Center of Materials Science and Optoelectronics Engineering \& CAS Center of Excellence in Topological Quantum Computation, University of Chinese Academy of Sciences, Beijing 100049, China}
\address[c]{Shenzhen Engineering Center for the Fabrication of Two-Dimensional Atomic Crystals, Shenzhen Institutes of Advanced Technology, Chinese Academy of Sciences, Shenzhen 518055, P. R. China}
\address[d]{College of Physics Science and Technology, Hebei University, Baoding 071002, China}
\address[e]{Beijing Academy of Quantum Information Science, Beijing 100193, China}
\fntext[1]{These authors contributed equally to this work}

\begin{abstract}
The contents of the Supplementary Information are summarized as follows: (1) complex Refractive indexes of black phosphorus along three principle axes; (2) the calculation of ARPR intensity in black phosphorus at normal laser incidence; (3) the amplitude transmission coefficient of $s$- and $p$-polarization components at the air/BP interface; (4) the calculation of ARPR intensity in black phosphorus at oblique laser incidence; (5)Fitted Raman tensors of bulk BP by the BLD and PDB models at 488 nm and 532 nm; (6) The calculated ARPR intensities of bulk BP at oblique laser incidence based on the PDB model.
\end{abstract}




\end{frontmatter}




\section{Complex Refractive indexes of black phosphorus along three principle axes}

\begin{figure}[!htb]
\centerline{\includegraphics[width=80mm,clip]{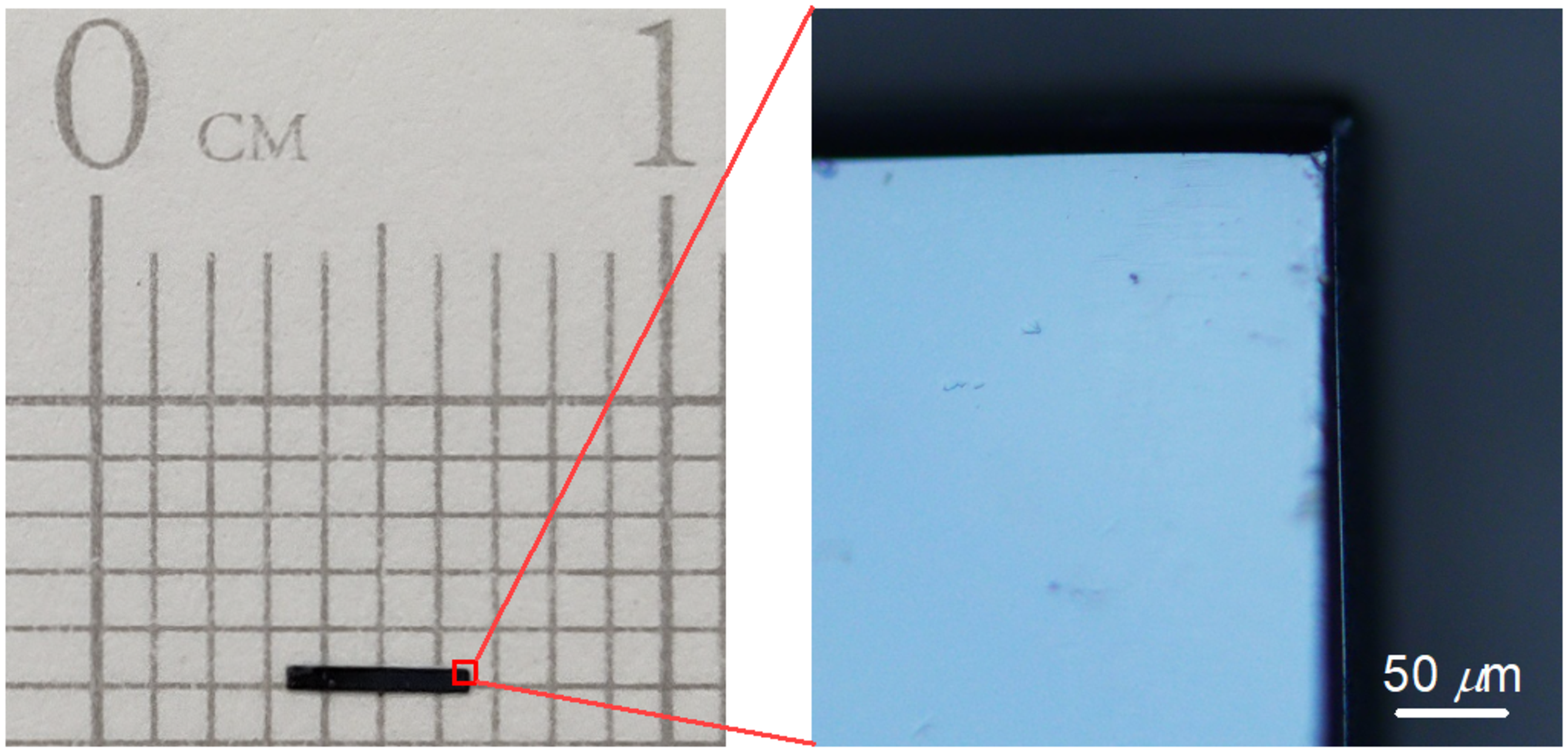}}
\caption{(Color online) Typical optical image of bulk BP.}\label{Fig-S1}
\end{figure}

Here, the complex refractive indexes along $x$, $y$ and $z$ axes were obtained by the incident-angle ($\delta_{\rm i}$) dependent reflectance with $\delta_{\rm i}$ ranging from 0$^{\circ}$ to 70$^{\circ}$ according to the Fresnel$'$s law. Based on Fresnel$'$s law, if the light is traveling from medium $i$ to $j$, the amplitude reflection coefficient of $s$- ($r_{ij}^s$) and $p$-polarization ($r_{ij}^p$) components are
\begin{equation}
r_{ij}^s = \frac{\tilde{n}_{is} {\rm cos}\delta_{i} - \tilde{n}_{js} {\rm cos}\delta_j}{\tilde{n}_{is} {\rm cos}\delta_i + \tilde{n}_{js} cos\delta_j}
\end{equation}
\begin{equation}
r_{ij}^p = \frac{\tilde{n}_{jp} {\rm cos}\delta_i - \tilde{n}_{ip} {\rm cos}\delta_j}{\tilde{n}_{jp} {\rm cos}\delta_i + \tilde{n}_{ip} {\rm cos}\delta_j}
\end{equation}

\noindent respectively, where $\delta_i$ and $\delta_j$ are the incident and refraction angle, respectively. $\tilde{n}_{is}$ ($\tilde{n}_{ip}$) and $\tilde{n}_{js}$ ($\tilde{n}_{jp}$) represent the complex refractive index of medium $i$ (air) and $j$ (BP) for $s$-polarization ($p$-polarization), respectively. Accordingly, the intensity reflectivity of the $s$- ($R_s$) and $p$-polarized ($R_p$) light at the surface between air and BP sample are $R_s=|r_{ij}^s|^2$ and $R_p=|r_{ij}^p|^2$, respectively. The high quality BP sample (Fig. \ref{Fig-S1}) make it possible to measure its $R_s$ and $R_p$ by the experimental setup shown in Fig.2(a). The bulk BP sample is thick enough and the substrate effect can be neglected here. For the air, $\tilde{n}_{is}$=$\tilde{n}_{ip}$ due to the isotropy. Thus, we should only consider the varied refractive index for $s$- and $p$-polarization in BP sample. For simplicity, here we neglect the subscript $j$ for the refractive index of $s$- and $p$-polarization components in BP sample. As shown in Fig.2(a), for the in-plane configuration, only the $s$-polarization along the rotation axis ($i.e.,$ $x$ or $z$ axis) exists in the BP sample, $i.e.,$ $\tilde{n}_s=\tilde{n}_x$ or $\tilde{n}_s=\tilde{n}_z$. Therefore, by fitting the $\delta_{\rm i}$-dependent intensity reflectivity ($R_s$) of a specific laser in in-plane configuration for $x$- and $z$-axis rotation geometries, the complex refractive index of the $x$ and $z$ at the corresponding wavelength can be obtained, respectively. For the out-of-plane configuration, only the $p$-polarization exist, exhibiting an angle of $\delta_{\rm i}$ relative to the $z$ ($x$) axis for $x$($z$)-axis geometry. The complex refractive index of $p$-polarization $\tilde{n}_p$ is expressed as $\tilde{n}^2_p = \frac{\tilde{n}_{\rm para}^2 \tilde{n}_{\rm perp}^2}{\tilde{n}_{\rm para}^2 {\rm sin}^2\delta_j + \tilde{n}_{\rm perp}^2 {\rm cos}^2\delta_j}$, where $\tilde{n}_{\rm perp}$ and $\tilde{n}_{\rm para}$ represent the refractive index for electric field polarized perpendicular ($\tilde{n}_y$) and parallel ($\tilde{n}_z$ or $\tilde{n}_x$) to the basal plane of BP sample, respectively. Thus, the laser intensity reflectivity in out-of-plane configuration ($p$-polarization) is related to $\tilde{n}_y$, $\tilde{n}_z$ and $\tilde{n}_y$, $\tilde{n}_x$ in $x$- and $z$-axis geometries, respectively. Thereby, the corresponding complex refractive indexes for a given wavelength can be obtained by fitting the intensity reflectivity to the Fresnel$'$s law and the fitted $\tilde{n}$ in $x$, $y$ and $z$ axes are listed in Table 1 and Table \ref{tb-S1}.

\begin{table}[htbp]
\centering\caption{Complex refractive indexes $\tilde{n}$ along three principle axes of bulk BP at 488 nm and 532 nm.}
\begin{tabular}{cccc}
\hline
Wavelength & $\tilde{n}_x$ & $\tilde{n}_y$ & $\tilde{n}_z$\\
\hline
488 nm & 4.82+0.067i & 4.62+0.73i & 4.72+0.71i\\
532 nm & 4.25+0.054i & 4.06+0.32i & 4.10+0.55i\\
\hline
\end{tabular}
\label{tb-S1}
\end{table}


\section{The calculation of ARPR intensity in black phosphorus at normal laser incidence}

When the linearly polarized laser is normally incident onto the surface (basal plane) of BP, the $s$- and $p$-polarization are along $z$ and $x$ axes, respectively. Thus, the electric field of incident light can be decomposed into two components along the zigzag ($x$) and armchair ($z$) directions. Because the polarization of incident and scattered light was changed by varying the angle ($\theta$/2) between the fast axis of the half-wave plate and $x$ axis of BP, here we introduce the Jones matrix of the half-wave plate in $x$-$z$ plane\cite{Jones-JOSA-1941},
\begin{equation}
J={
\left( \begin{array}{ccc}
{\rm cos}\theta & 0 & -{\rm sin}\theta\\
0 & 0 & 0\\
{\rm sin}\theta & 0 & {\rm cos}\theta
\end{array}
\right )},
\end{equation}
The polarization of the incident laser, $e_{\rm i0} = \left(1,0,0\right)$, would be changed after passing through the half-wave plate, $i.e.,$ $e_{\rm i} = J \cdot e_{\rm i0} = \left({\rm cos}\theta,0,{\rm sin}\theta\right)$\cite{Liu-cpb-2017}. Thus, $E^x_{\rm i0} = E_{\rm i0} {\rm cos}\theta$, $E^z_{\rm i0} = E_{\rm i0} {\rm sin}\theta$, where $E_{\rm i0}$ is the intensity of the electric field at the surface of the sample. As known, the electric field propagating within the medium would satisfies,
\begin{equation}
E = E_0 {\rm e}^{2 \uppi i\tilde{n} Y / \lambda}
\end{equation}
where $E_0$ is the initial electric field, $\tilde{n}$ is the complex refractive index of the medium, $Y$ is the propagation distance and $\lambda$ is the wavelength of the light. In anisotropic bulk BP sample, $E^x_{\rm i0}$ and $E^z_{\rm i0}$ experience different phase shifts and penetration depths due to the varied complex refractive indexes along $x$ and $z$ axes,
\begin{equation}
\begin{split}
& E^x_{{\rm i}Y} = E_{\rm i0} {\rm e}^{2 \uppi i\tilde{n}_x Y / \lambda} {\rm cos}\theta=E_{\rm i0} e^x_{{\rm i}Y},\\
& E^z_{{\rm i}Y} = E_{\rm i0} {\rm e}^{2 \uppi i\tilde{n}_z Y / \lambda} {\rm sin}\theta=E_{\rm i0} e^z_{{\rm i}Y}
\end{split}
\end{equation}
\noindent where $\tilde{n}_x$ and $\tilde{n}_z$ are the complex refractive indexes of BP along the $x$ and $z$ axes, respectively. The electric field components of the Raman signal occurring at site $Y$ ($E^u_{\rm r0}$) is associated with the Raman tensor and $E^v_{{\rm i}Y}$ ($v=x,y,z$),
\begin{equation}
E^u_{\rm r0}=\sum_{v=x,y,z}R_{uv}E^v_{{\rm i}Y}, u=x,y,z
\end{equation}
\noindent The electric field components of scattered light also experience the depth-dependent polarization and intensity within the sample. In addition, the polarization of the scattered light is also changed by half-wave plate and only the electric field components parallel to that of the incident laser can be collected by the spectrometer for the geometry with parallel polarization configuration. Thus, the electric field components along $x$ and $z$ axes of collected Raman signal are
\begin{equation}
\begin{split}
& E^x_{{\rm r}Y} =E^x_{\rm r0} {\rm e}^{2 \uppi i\tilde{n}_x Y / \lambda} {\rm cos}\theta=E^x_{\rm r0} e^x_{{\rm r}Y},\\
& E^z_{{\rm r}Y} =E^z_{\rm r0} {\rm e}^{2 \uppi i\tilde{n}_z Y / \lambda} {\rm sin}\theta=E^z_{\rm r0} e^z_{{\rm r}Y}
\end{split}
\end{equation}
\noindent respectively, where $e^{x(z)}_{{\rm i}Y}$=$e^{x(z)}_{{\rm r}Y}$. Here we ignore the differences in wavelength and $\tilde{n}$ between the scattered light and incident light due to the small phonon energy. The scattered signal intensity is obtained by summing the squares of electric field components integrating over the penetration depth along respective crystallographic orientation,
\begin{equation}
\label{eq9}
\begin{split}
& I\propto\sum_{u=x,y,z}\left|\int_{0}^{\infty}E^u_{{\rm r}Y}dY\right|^2,\\
& \propto\sum_{u=x,y,z}\left|\int_{0}^{\infty}e^u_{{\rm r}Y}\cdot\sum_{v=x,y,z}R_{uv}\cdot e^v_{{\rm i}Y}dY\right|^2
\end{split}
\end{equation}
\noindent Once the real Raman tensor and complex refractive indexes along principle axes are known, the ARPR intensity of a specific Raman mode under a given wavelength excitation can be anticipated.

\section{The amplitude transmission coefficient of $s$- and $p$-polarization components at the air/BP interface}
When the laser light is incident onto the basal plane of BP with an incident angle of $\delta_{\rm i}$, the $\delta_{\rm i}$-dependent amplitude transmission coefficients should be considered for both $s$- and $p$-polarization components. According to the Fresnel$'$s law, the amplitude transmission coefficients of $s$- and $p$-polarization components are
\begin{equation}
\begin{split}
& t_{ij}^s = \frac{2 \tilde{n}_{is} {\rm cos}\delta_{\rm i}}{\tilde{n}_{is} {\rm cos}\delta_{\rm i} + \tilde{n}_{js} {\rm cos}\delta_j},\\
& t_{ij}^p = \frac{2 \tilde{n}_{ip} {\rm cos}\delta_{\rm i}}{\tilde{n}_{jp} {\rm cos}\delta_{\rm i} + \tilde{n}_{ip} {\rm cos}\delta_j}
\end{split}
\label{tsp}
\end{equation}
respectively.

\section{The calculation of ARPR intensity in black phosphorus at oblique laser incidence}

Based on the experimental setup shown in Fig.2(a), the $s$-polarization and $p$-polarization can be dealt with separately. We take the sample rotating around the $x$ axis as an example. For in-plane configuration, only the electric field component in the $x$ direction exists in BP sample, $i.e.,$ $\tilde{n}_s$=$\tilde{n}_x$. Considering $t_{ij}^s$ at the air/BP interface,
\begin{equation}
E_{\rm i0}^{s} = t_{ij}^{s} E_{\rm i0},
\end{equation}
\noindent where $E_{\rm i0}$ is the electric field at the surface and $t_{ij}^{s}$ can be calculated by Eq.\ref{tsp}. Within the sample, the electric field of incident ($E_{\rm i0}^{s}$) and scattered light ($E_{\rm r0}^{s}=\sum_{v=x,y,z}R_{sv}E^v_{{\rm i}Y}$) propagating a distance of $Y$ would become,
\begin{equation}
\begin{split}
& E_{{\rm i}Y}^{s} = E_{\rm i0}^{s} {\rm e}^{2 \uppi i\tilde{n}_{s} Y / \lambda},\\
& E_{{\rm r}Y0}^{s} = E_{\rm r0}^{s} {\rm e}^{2 \uppi i\tilde{n}_{s} Y / \lambda}
\end{split}
\end{equation}
\noindent Thus, the electric field of the scattered light that passes through the BP/air surface is
\begin{equation}
E_{{\rm r}Y}^{s} = t_{ji}^{s}E_{\rm r0}^{s} {\rm e}^{2 \uppi i\tilde{n}_{s} Y / \lambda}
\end{equation}
Therefore, the polarized Raman scattering for in-plane configuration can be calculated by Eq.\ref{eq9}.

As for the out-of-plane configuration, there is only electric field of $p$-polarization,
\begin{equation}
E_{\rm i0}^{p} = t_{ij}^{p} E_{\rm i0},
\end{equation}
\noindent where $t_{ij}^p$ can be calculated by Eq.\ref{tsp} with $\tilde{n}^2_p = \frac{\tilde{n}_{\rm para}^2 \tilde{n}_{\rm perp}^2}{\tilde{n}_{\rm para}^2 {\rm sin}^2\delta_j + \tilde{n}_{\rm perp}^2 {\rm cos}^2\delta_j}$ ($\tilde{n}_{\rm perp}$=$\tilde{n}_y$, $\tilde{n}_{\rm para}$=$\tilde{n}_z$). After propagating a distance of $Y$, $E_{{\rm i}Y}^{p}$ would provide electric field components along the $y$ and $z$ directions, $i.e.,$ $E_{{\rm i}Y}^{yp}$ and $E_{{\rm i}Y}^{zp}$,
\begin{equation}
E_{{\rm i}Y}^{yp} = E_{\rm i0}^{p} {\rm e}^{2 \uppi i\tilde{n}_{p} Y / \lambda}{\rm sin}\delta_j,
E_{{\rm i}Y}^{zp} = E_{\rm i0}^{p} {\rm e}^{2 \uppi i\tilde{n}_{p} Y / \lambda}{\rm cos}\delta_j,
\end{equation}
For the electric field components of scattered light, $i.e.,$ $E_{\rm r0}^{y(z)p} =\sum_{v=x,y,z}R_{y(z)v}E^v_{{\rm i}Y}$, only those along the $p$-polarization can be collected by the spectrometer. Considering the propagation distance of $Y$, these electric field components are,
\begin{equation}
 E_{{\rm r}Y0}^{yp} = E_{\rm r0}^{yp} {\rm e}^{2 \uppi i\tilde{n}_{p} Y / \lambda}{\rm sin}\delta_j,
 E_{{\rm r}Y0}^{zp} = E_{\rm r0}^{zp} {\rm e}^{2 \uppi i\tilde{n}_{p} Y / \lambda}{\rm cos}\delta_j
\end{equation}
\noindent where $\delta_j$ is the refraction angle associated with $\delta_{\rm i}$. Accordingly, the electric field components of the scattered light that pass through the BP/air surface are
\begin{equation}
\begin{split}
& E_{{\rm r}Y}^{yp} = t_{ji}^pE_{\rm r0}^{yp} {\rm e}^{2 \uppi i\tilde{n}_{p} Y / \lambda}{\rm sin}\delta_j,\\
& E_{{\rm r}Y}^{zp} = t_{ji}^pE_{\rm r0}^{zp} {\rm e}^{2 \uppi i\tilde{n}_{p} Y / \lambda}{\rm cos}\delta_j
\end{split}
\end{equation}
\noindent Thus, $\delta_{\rm i}$-dependent Raman intensity in out-of-plane configuration can be anticipated based on Eq.\ref{eq9}. The corresponding experimental and calculated results excited by 532 nm and 488 nm for $x$- and $z$-axis rotation geometries are shown in Fig.4, which are in agreement with each other.

\vspace*{10mm}
\section{Fitted Raman tensors of bulk BP by the BLD and PDB models at 488 nm and 532 nm}

\begin{table*}[!htbp]
\centering
\caption{Fitted real Raman tensors in bulk BP including elements $a$, $b$ and $c$ by the BLD model proposed in the present work and fitted complex Raman tensors in bulk BP including elements ($a$, $b$ and $c$) and phase difference ($\phi_{ca}$ and $\phi_{ba}$) of the $A_{\rm g}^2$ and $A_{\rm g}^1$ modes by the PDB model at 488 nm and 532 nm. The absolute value of Raman tensor element $a$ is fixed at 1.00 and the confidence interval for other Raman tensor elements are given at a 95\% confidence level. in which n.a. means the term is not applicable.}
\resizebox{\textwidth}{15mm}{%
\begin{tabular}{cccccccccccccc}
\hline
\multicolumn{2}{c}{\multirow{2}{*}{Wavelength}}&
\multicolumn{2}{c}{Raman mode} &\,&
\multicolumn{4}{c}{$A_{\rm g}^2$}&\,&\multicolumn{4}{c}{$A_{\rm g}^1$} \\
\cline{3-4}\cline{6-9}\cline{11-14}
\multicolumn{2}{c}{}&
\multicolumn{2}{c}{Raman tensor}&\,&a&b&c&$\phi_{ca}~/~\phi_{ba}$&\,&a&b&c&$\phi_{ca}~/~\phi_{ba}$\\
\hline
\multicolumn{2}{c}{\multirow{2}{*}{488nm}}&
\multicolumn{2}{c}{BLD}&\,&{ 1.00}&${0.81\pm0.20}$&${2.70\pm0.28}$&n.a.&\,&{1.00}&${0.53\pm0.13}$&${1.05\pm0.13}$&n.a.\\
\cline{3-14}
\multicolumn{2}{c}{}&
\multicolumn{2}{c}{PDB}&
\,&1.00&$0.06\pm0.04$&$1.79\pm0.13$&$84.3^{\circ}/0.02^{\circ}$&
\,&1.00&$0.14\pm0.06$&$0.26\pm0.03$&$0.01^{\circ}/0.17^{\circ}$\\
\hline
\multicolumn{2}{c}{\multirow{2}{*}{532nm}}&
\multicolumn{2}{c}{BLD}&\,&{1.00}&${2.40\pm0.11}$&${0.56\pm0.05}$&n.a.&\,&{1.00}&${2.40\pm0.11}$&${0.42\pm}$n.a.&n.a.\\
\cline{3-14}
\multicolumn{2}{c}{}&
\multicolumn{2}{c}{PDB}&
\,&1.00&$1.45\pm0.05$&$0.84\pm0.03$&$88.3^{\circ}/93.2^{\circ}$&
\,&1.00&$1.46\pm0.04$&$0.44\pm0.02$&$0.01^{\circ}/79.7^{\circ}$\\
\hline
\end{tabular}}
\label{tb-S2}
\end{table*}

\section{The calculated ARPR intensities of bulk BP at oblique laser incidence based on the PDB model.}

\begin{figure*}[!htb]
\centerline{\includegraphics[width=140mm,clip]{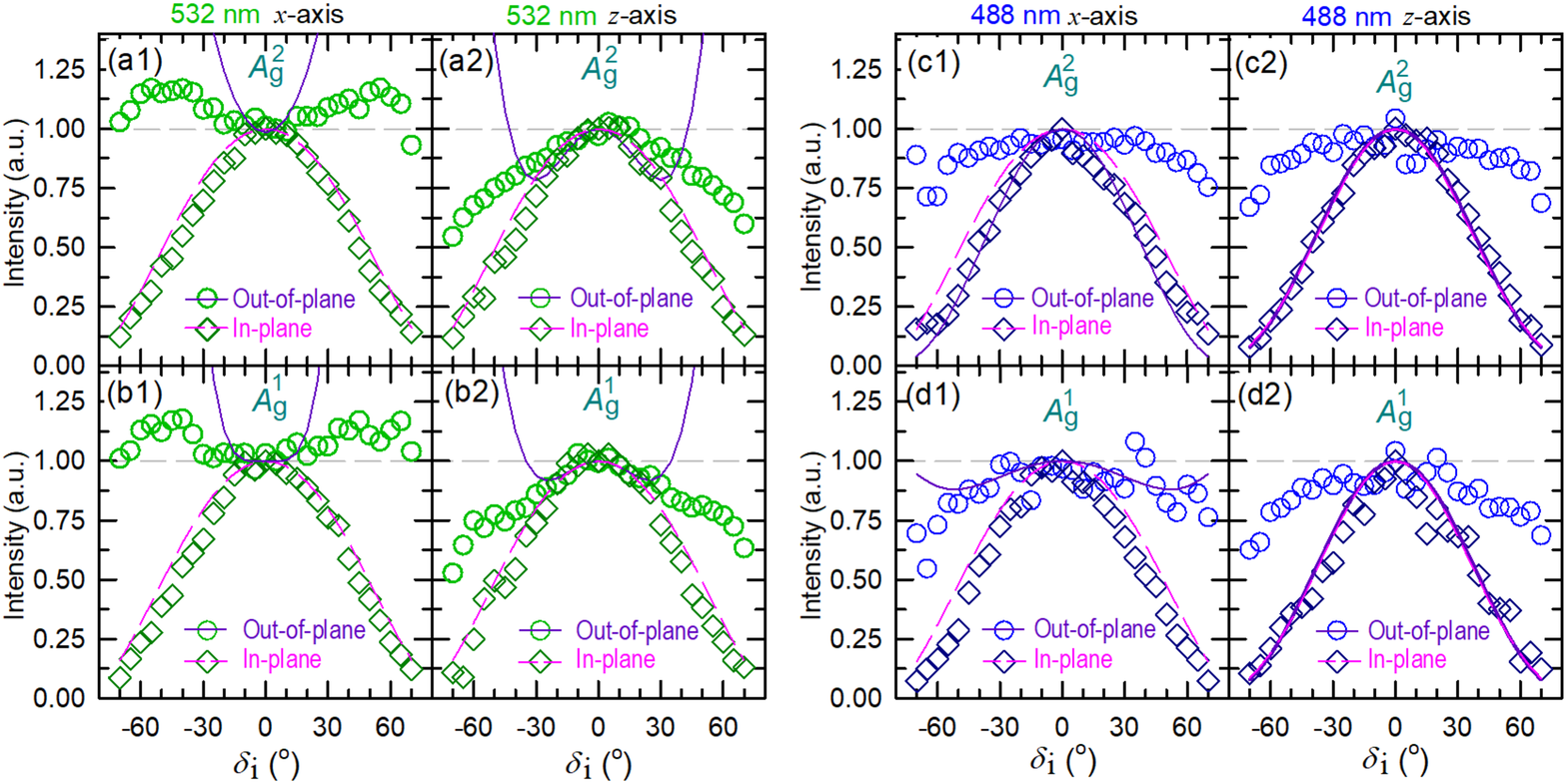}}
\caption{(Color online) ARPR spectra of bulk BP at oblique laser incidence. The experimental (open diamonds and circles) and calculated (dashed and solid lines) $\delta_{\rm i}$-dependent ARPR intensity of the $A_{\rm g}^2$ and $A_{\rm g}^1$ modes excited by 532 nm laser under in-plane and out-of-plane configurations in $x$- (a1,b1) and $z$-axis (a2,b2) rotation geometries, respectively. The corresponding results excited by 488 nm laser are shown in (c,d). The calculated ARPR intensities are based on the PDB model with $\delta_{\rm i}$-dependent reflectivity considered.}\label{Fig-S2}
\end{figure*}

\bibliographystyle{csb}
\bibliography{BP-ref}
